\documentclass[preprint,notoc]{JHEP3} % 10pt is ignored!
\usepackage{epsfig,multicol}
\usepackage{amssymb}
\usepackage{graphics}

\usepackage{amsmath,latexsym}
\def\beq{\begin{eqnarray}}    %%%  begequation/eqnarray
\def\eeq{\end{eqnarray}}      %%%  endequation/eqnarray

\usepackage{amssymb}
\usepackage{amsmath}
\usepackage{graphicx}

\renewcommand{\vec}[1]{{\bf #1}}

\def\beq{\begin{eqnarray}}
\def\eeq{\end{eqnarray}}
\def\ln{\,\mbox{ln}\,}

\def\CC{\rho_\Lambda}

\def\de{\delta}

\def\na{\nabla}
\def\pa{\partial}

\def\th{\theta}

\def\Ga{\Gamma}

\def\La{\Lambda}

\def\Om{\Omega}

\def\hh{\hat{h}}
\title{Density Perturbations for Running
 Cosmological Constant}
\author{
J\'ulio C. Fabris $^{a}$, Ilya L. Shapiro $^{b}$, Joan Sol\`{a}
$^{c,d}$
\\
$^{a}\,$Departamento de F{\'\i}sica -- CCE, Universidade Federal
do Esp{\'\i}rito Santo, ES, Brazil,
\\
E-mail: fabris@cce.ufes.br
\\
$^{b}\,$Departamento de F\'{\i}sica - Instituto de Ci\^encias
Exatas
\\
Universidade Federal de Juiz de Fora, 36036-330, MG, Brazil
\\
E-mail: shapiro@fisica.ufjf.br
\\
$^{c}$ High Energy Physics Group, Dep. E.C.M.  Universitat de
Barcelona,
\\
\,Diagonal 647, Barcelona, Catalonia, Spain \vskip 2mm $^{\,d}\,$
 C.E.R. for Astrophysics, Particle Physics and Cosmology
\footnote{Associated with Instituto de Ciencias del Espacio-CSIC.}
\\
E-mail: sola@ifae.es.
}

\preprint{ UB-ECM-PF-06/14}

\abstract{The dynamics of density and metric perturbations is
investigated for the previously developed model where the decay of
the vacuum energy into matter (or vice versa) is due to the
renormalization group (RG) running of the cosmological constant (CC)
term. The evolution of the CC depends on the single parameter $\nu$,
which characterizes the running of the CC produced by the quantum
effects of matter fields of the unknown high energy theory below the
Planck scale. The sign of $\nu$ indicates whether bosons or fermions
dominate in the running.  The spectrum of perturbations is computed
assuming an adiabatic regime and an isotropic stress tensor.
Moreover, the perturbations of the CC term are generated from the
simplest covariant form suggested by the RG model under
consideration. The corresponding numerical analysis shows that for
$\nu>0$ there is a depletion of the matter power spectrum at low
scales (large wave numbers) as compared to the standard $\Lambda$CDM
model, whereas for $\nu<0$ there is an excess of power at low
scales. We find that the LSS data rule out the range
$\left|\nu\right|> 10^{-4}$ while the values $\left|\nu\right|\leq
10^{-6}$ look perfectly acceptable. For $\nu<0$ the excess of power
at low scales grows rapidly and the bound is more severe. From the
particle physics viewpoint, the values \ $|\nu|\simeq 10^{-6}$
correspond to the ``desert'' in the mass spectrum above the GUT
scale $M_X\sim 10^{16}\,GeV$. Our results are consistent with those
obtained in other dynamical models admitting an interaction between
dark matter and dark energy. We find that the matter power spectrum
analysis is a highly efficient method to discover a possible scale
dependence of the vacuum energy.}


\begin{document}
%%%%%%%%%%%%%%%%%%%%%%%%%%%%%%%%%%%%%%%%%%%%%%%%%%%%%%%%%%%
%%%%%%%%%%%%%%%%%%%%%%%%%%%%%%%%%%%%%%%%%%%%%%%%%%%%%%%%%%%

\section{Introduction}
\label{sect:Intro}

The analysis of cosmic perturbations\,\cite{peebles,LL,weinberg}
represents one of the main tests for the candidate cosmological
models and especially for the models of a time-dependent Dark
Energy (DE). The recent data on the accelerated expansion of the
Universe from the high redshift Type Ia supernovae
\cite{Supernovae}, combined with the measurements of the CMB
anisotropies \cite{CMBR,WMAP, WMAP3Y} and
LSS\,\cite{cole,tegmark} indicate that the DE is responsible for
most of the energy balance in the present Universe. The natural
candidate to be the DE is the cosmological constant, mainly if we
take into account that from the quantum field theory (QFT) point
of view the zero value of the CC would be extremely
unnatural\,\cite{weinRMP,nova,cosm}. The solution of the standard
CC problems, such as the origin of the fine-tuning between vacuum
and induced counterparts and the problem of coincidence, is not
known. Historically the first attempts to solve some of these
problems went along the lines of dynamical adjustments
mechanisms\,\cite{Dolgov}. Nowadays the discussion is approached
from many different perspectives and has generated an extensive
literature, see e.g. \,\cite{CCRev,Copeland06} and references
therein. One of the most interesting aspects of these discussions
is the possibility to have a slowly varying DE density of the
Universe\,\cite{RatraPeebles}. This option cannot be ruled out by
the analysis of the present data\,\cite{Alam,Jassal}. In case
that the new generation of cosmic observational experiments will
indeed detect a variable DE, the natural question to ask is:
would this fact really mean the manifestation of a qualitatively
new physical reality such as quintessence \cite{quint}, Chaplygin
gas \cite{chap}, extra dimensions \cite{extra} or of low-energy
quantum gravity \cite{PVT,pol,antmot,reuter}?  Let us emphasize
here that one cannot honestly address this question without also
exploring the possibility of a time-variable CC whose running is
due to known physical effects, that is, due to the quantum effects
of matter fields on a curved classical background\,\cite{book}.

The most natural manner to investigate the possible running of the
CC produced by the quantum effects of matter fields is the
renormalization group (RG) in curved space-time
\cite{lam,nova,cosm}. Unfortunately, the existing calculational
methods are not sufficient for performing a complete theoretical
investigation of this problem \cite{apco}. At the same time, the
phenomenological proposal \cite{nova,cosm,RGCC,IRGA03} shows that a
positive result can not be ruled out. The hypothesis of a standard
quadratic form of decoupling for the quantum effects of matter
fields at low energies \cite{nova,babic} leads to consistent
cosmological models of a running CC with  potentially observable
consequences \cite{CCfit,AHEP03,Gruni,SS1,SolSte}. Among the many
possibilities one can distinguish essentially two different models
of running CC in QFT in curved space-time \footnote{ Generalizations
of this kind of semiclassical RG cosmological models (which may e.g.
help to alleviate the cosmic coincidence problem) are also possible,
see \cite{LXCDM1}}. The first one \cite{RGCC,IRGA03,CCfit} admits
that there is energy exchange between matter and vacuum sectors of
the theory\,\footnote{See \,\cite{BarrowClifton} for a general
discussion of energy exchange between cosmic fluids.}. This leads to
the vacuum energy density function $\rho_\La=\rho_\La(H)$ associated
to the cosmological term, $\Lambda$, while the Newton constant $G$
remains invariable\,\footnote{Let us notice that here we follow
\cite{nova} and identify the cosmic energy scale with the Hubble
parameter $H$, despite other choices are also possible \cite{nova,
cosm, reuter,babic,baier}.}. The second model does not admit a
matter-vacuum energy exchange, but has both parameters
scale-dependent $G=G(H)$ and $\rho_\La=\rho_\La(H)$\,\cite{Gruni}.
In the present work we shall concentrate on the analysis of the
perturbations for the first model only, and postpone the
investigation of the second model for another publication. Let us
remark that the analysis of cosmic perturbations in the first model
has already been addressed in \cite{OphPel} using an indirect
procedure based on bounding the amplification of the density matter
spectrum in the recombination era caused by vacuum decay into CDM.
This amplification is necessary to compensate for the dilution of
the $\delta\rho_m/\rho_m$ matter spectrum at low redshifts resulting
from the enhanced matter-radiation density $\rho_m$ -- associated to
the decay of the vacuum into CDM. A qualitatively similar analysis
has been performed in \cite{WangMeng}, and also in \cite{AlcLim}
using the so-called CMB shift parameter\,\cite{WMAP}. It is worth
noticing that the upper bounds for the running obtained in all these
references disagree between themselves, in some cases at the level
of several orders of magnitude. This severe divergence illustrates
the importance and necessity of performing a direct calculation of
the perturbations, which we undertake here thoroughly. For example,
the estimate used in \cite{OphPel} was derived from the
phenomenological study presented in Ref.\,\cite{Peacock}, where the
matter power spectrum of density fluctuations measured by the 2dF
Galaxy Redshift Survey (2dFGRS) was compared with the corresponding
matter power spectrum derived from the measurements of the CMB
anisotropies. The comparison served to place a bound at the level of
$10\%$ on the maximum possible difference between the two
derivations of the matter power spectrum\,\cite{Peacock}. By
requiring that the amplification of the density fluctuations of the
RG model at the recombination era do not surpass this bound the
authors of \cite{OphPel} were able to obtain a corresponding bound
on the the RG parameter of\,\cite{RGCC}, specifically $\nu<10^{-3}$.
Of course this bound is only approximate as it does not come from a
direct calculation of the density perturbations in the RG model.
Notwithstanding, as we shall see in what follows, there is a
qualitative agreement between these results and the ones obtained in
the direct calculation that we present here. In both cases the RG
models with Planck-scale mass particles are not favored and one is
led to conclude that a significant energy gap (or ``desert'') below
the Planck scale, $M_P$, is necessary to avoid contradictions
between the calculated spectrum of the density perturbations and the
observational data. The favored mass scale turns out to be three
orders of magnitude below $M_P$, and can be identified with the
typical GUT scale $M_X\sim 10^{16}\,GeV$. In this paper we will
provide a rigorous quantitative derivation of this result. Let us
also remark that the analysis of perturbations presented here is not
covered by the many works in the literature dealing with
perturbations in a Universe with dynamical dark energy -- see e.g.
\,\cite{Copeland06} and references therein. In fact, the nature of
the DE under consideration is not directly related to the properties
of dynamical scalar field(s), but to the QFT running of the
cosmological constant. Therefore, our approach is qualitatively new
and has not been dealt with in the literature before. It
constitutes, to our knowledge, the first computation of density
perturbations in the presence of a running cosmological term. This
is important in order to assess the impact on structure formation
from a possible scale dependence of the vacuum energy.
\\ \indent The paper is organized as follows. In the next section
we summarize the necessary information about the variable CC
model under discussion. The reader may consult the papers
\cite{nova,RGCC,CCfit,AHEP03} for further technical details and
extensive discussions. In section 3 we derive the equations for
the perturbations. Section 4 is devoted to the numerical solution
of these equations and to their comparison with the
2dFGRS\,\cite{cole}. Finally, in the last section we draw our
conclusions.

%%%%%%%%%%%%%%%%%%%%%%%%%%%%%%%%%%%%%%%%%%%%%%%%%%%%%%%
%%%%%%%%%%%%%%%%%%%%%%%%%%%%%%%%%%%%%%%%%%%%%%%%%%%%%%%
\section{The background cosmological solution in the running CC model}
\label{sect:VCT}

In the framework under consideration\cite{nova,RGCC}, with
renormalization group corrected CC and energy exchange between
vacuum and matter sectors, the cosmological evolution is governed
by the following three ingredients: \ $(i)$ Friedmann equation,
\beq H^2(z)=\frac{8\,\pi\,G}{3}\, \big[\rho_m(z)+\rho_\La(z)\big]
\,+\,H_0^2\Omega_k^0\,(1+z)^2\,, \label{Friedmann} \eeq where \
$\rho_m(z)$ \ and \ $\rho_\La(z)$ \ are the densities of matter
and vacuum energies taken as functions of the cosmological
redshift, $z$, with  \ $H_0^2\,\Omega_k^0(1+z)^2=-k/a^2$; \
$(ii)$ Conservation law (following from the Bianchi identity)
which regulates the exchange of energy between matter and vacuum,
\beq
\frac{d\rho_\La}{dz}+\frac{d\rho_m}{dz}\,=\,\frac{3\,\rho_m}{1+z}\,,
\label{conservation} \eeq and \ $(iii)$ the renormalization group
equation for the CC density, assuming a soft form of decoupling
for the cosmological constant \cite{nova,RGCC,babic} \footnote{For
a dimensionless parameter this would be ${\cal O}(p^2/m^2)$ (that
is the standard Appelquist and Carazzone \cite{AC}) decoupling
law. In the special case of vacuum energy, which is a dimension-4
parameter, this leads to the quadratic dependence on the scale,
which we associate to the Hubble parameter\,\cite{nova}.
Inclusion of bulk viscosity effects would lead to linear terms in
$H$\,\cite{bulkvisco}.  We neglect them in the present work.} \beq
\frac{d\rho_\La}{d{\ln H}}= \frac{1}{(4\pi)^{2}}\
\sigma\,H^{2}M^{2}\,. \label{RG} \eeq In this equation, the
effective mass\ $M$ \ represents an additive sum of the
contributions of all virtual massive particles, and $\,\sigma=\pm
1$ depending on whether fermions or bosons dominate at the
highest energies. It proves useful introducing the new
(dimensionless) parameter \beq
 \label{nu} \nu \equiv
\frac{\sigma}{12\,\pi}\,\frac{M^2}{M_P^2}\,.
\eeq
The value of $\nu$ defines the strength of the quantum effects and
has direct physical interpretation. For example, taking the
``canonical'' value $\nu=\nu_0\equiv 1/12\pi= 0.026$ means that
the effective cumulative mass $M$ of all virtual massive
particles is equal to the Planck mass, whereas $|\nu|\sim 1$
means the existence of a particle with trans-planckian mass or of
nearly Planck-mass particles with huge multiplicities. Taking
$|\nu|\sim 10^{-6}$ means that the spectrum of particles is
bounded from above at the (GUT) scale \ $M_X\approx 10^{16}\,GeV
\sim 10^{-3}M_P$. Much smaller values of \ $|\nu|\ll 10^{-6}$ \
mean the existence of an extra unbroken symmetry between bosons
and fermions (e.g. supersymmetry) at the GUT scale etc.

We already noticed that Eq.\,(\ref{conservation}) implies the
possibility of an energy transfer between vacuum and matter,
which may be interpreted as a decay of the vacuum into matter or
vice versa. While the decay of the vacuum energy into ordinary
matter and radiation could be problematic for the observed CMB,
the decay into CDM can be allowed\,\cite{RatraPeebles,OpherPel}
provided the rate is sufficiently small. It is of course part of
the present study to evaluate the possible size of this rate.

The complete analytic solution of the equations (\ref{Friedmann}),
(\ref{conservation}), (\ref{RG}) can be found in
\cite{RGCC,CCfit}. Here we present only those formulas which are
relevant for the perturbations calculus. First of all, from
(\ref{RG}) it follows that
\beq
\rho_\La(z;\nu)\,=\,\rho_\Lambda^0 + \frac{3\,\nu}{8\pi}\,M_P^2\,
\left(H^2-H^2_0\right)\,.
\label{Friedmann2}
\eeq
Furthermore, we
introduce notations for the pressure and density of the vacuum
and matter components, such that the total values become (we
assume pressureless matter at present)
\beq
\rho_t=\rho_m+\rho_\La\,,\qquad P_t=P_m+P_\La\,, \qquad
\mbox{where}\qquad P_\La=-\rho_\La\,, \qquad P_m=0\,.
\label{total}
\eeq
%%%%%%%%%%%%%%%%%%%%%%%%%%%%%%%%%%%%%%%%%%%%%%%%

The quantities (\ref{total}) determine the dynamical evolution law for
the scale factor:
\beq
{\dot H}+H^2= \frac{{\ddot a}}{a}=-\frac{4\pi
G}{3}\,(\rho_t + 3P_t) =-\frac{4\pi G}{3}\,(\rho_m - 2\rho_\La)\,.
\label{acceleration}
\eeq
What we shall need in what follows are the two ratios
\beq
f_1 &=& f_1(z)
\,\,=\,\, \frac{\rho_m}{\rho_t}   = \frac{(1+z)(H^2)^\prime
- 2H_0^2\,\Omega_k^0(1+z)^2} {3\left[H^2 -
H_0^2\,\Omega_k^0(1+z)^2 \right]}\,,
\nonumber
\\
\nonumber
\\
f_2 &=& f_2(z) \,\,=\,\, \frac{\rho_\La}{\rho_t} = \frac{3H^2
-(1+z)(H^2)^\prime - H_0^2\,\Omega_k^0(1+z)^2} {3\left[H^2 -
H_0^2\,\Omega_k^0(1+z)^2 \right]}\,, \label{ratios} \eeq where the
prime indicates  \ $d/dz$. The expansion rate \ $H(z)$ in this model
is given as follows: \beq H^2(z;\nu)\,=\,H_0^2 + \left(\Omega_M^0 -
\frac{2\nu\Omega_k^0}{1-3\,\nu}\right)\,
\frac{H_0^2\,\left[(1+z)^{3-3\nu}-1 \right]}{1-\nu} +
\frac{\Omega_k^0 H_0^2 (z^2+2z)}{1-3\,\nu} \,. \label{Hz} \eeq In
the last formula $\Omega_M^0=\Omega_{DM}^0+\Omega_B^0$ is the sum of
the Dark Matter (DM) and baryonic matter densities (relative to the
critical density) in the present-day Universe. Let us remark that
the scale dependence of the vacuum energy (\ref{Friedmann2}) does
not imply a change of the equation of state for the ``cosmological
constant fluid''. In this work we follow \cite{RGCC,IRGA03,CCfit}
and assume that \ $p_\La=-\rho_\La$, even when quantum corrections
are taken into account. It should be clear that this equation of
state is valid irrespective of whether $\Lambda$ is constant or
variable. A different issue is that the models with variable
cosmological parameters (like the present one) can be interpreted in
terms of an effective equation of state corresponding to a
self-conserved dark energy, as it is usually assumed in e.g. scalar
field models of the DE. The general framework for this effective EOS
analysis has been developed in \cite{SS1,SolSte}, where a particular
discussion of this issue for the present running CC model is also
included.

%%%%%%%%%%%%%%%%%%%%%%%%%%%%%%%%%%%%%%%%%%%%%%%%%%%%%%%%%%%%
%%%%%%%%%%%%%%%%%%%%%%%%%%%%%%%%%%%%%%%%%%%%%%%%%%%%%%%%%%%%
\section{Deriving the perturbations equations}
\label{sect:RGmodel1}

In order to derive the equations for the density and metric
perturbations we follow the standard formalism (see e.g.
\cite{peebles}-\cite{weinberg}). We will perform our calculation of
density perturbations under the following two assumptions:
\begin{itemize}
\item The entropic perturbations are negligible;
\item The full energy-momentum tensor is free of anisotropic
stresses.
\end{itemize}
In the absence of more information we believe that these hypotheses
are reasonable as they aim at the maximum simplicity in the
presentation of our results. Notice that in a single fluid case, the
pressureless fluid and the vacuum equation of state have zero
anisotropic pressure \cite{LL}; if both fluids interact, as it
happens in the present case, a contribution can appear, but we
cannot model it without introducing further free parameters.
Similarly, the contribution of entropic perturbations must be
introduced phenomenologically, and this would lead to the appearance
of more free parameters, thus masking any constraint we can obtain
on the fundamental parameter $\nu$ of our RG framework. It should
also be stressed that the same approximations are made in similar
works in the literature, for example in reference \cite{pavon} where
the density perturbations in the alternative framework of
interacting quintessence models are considered and a corresponding
bound on the parameter that couples dark energy and matter is
derived. Therefore, in order to better compare our results with
these (and other future) studies, and also to essentially free our
presentation of unnecessary complications extrinsic to the physics
of the RG running, in what follows we will adopt the aforementioned
set of canonical hypotheses. At the same time we compute the density
perturbations by neglecting the spatial curvature, which is
irrelevant at the epoch when these linear perturbations were
generated, as explained e.g. in \cite{weinberg}. Nevertheless we
have kept the spatial curvature in the background metric where it
could play a role at the present time.
\par
Let us first introduce the
4-vector velocity $U^\mu$. In the co-moving coordinates
$U^0=U_0=1$ and $U^i=U_i=0$. The total energy-momentum tensor of
matter and vacuum can be expressed via (\ref{total}) as
\beq
T_\mu^\nu = \big(\rho_t + P_t\,\big)\,U^\nu U_\mu -
P_t\,\de^\nu_\mu\,, \label{EMT}
\eeq
such that $T_0^0=\rho_t$ and
$T_i^j=-P_t\,\de_i^j$. It is pretty easy to derive the covariant
derivative of $U^\mu$:
\beq
 \na_\mu U^\mu  = 3H\,. \label{H}
 \eeq
The last equation is important, for it enables us to perturb the
Hubble parameter and eventually the running CC. In particular, using
(\ref{H}) we can rewrite Eq.\,(\ref{Friedmann2}) in the form
\beq
\rho_\La &=& {A} + {B}\,\big( \na_\mu U^\mu\big)^2\,, \label{A and
B}
\\
%% \label{lambda}
%% \eeq
%% \beq
\mbox{where}\qquad {A} &=& \rho_\La^0 -
\frac{3\,\nu}{8\pi}\,M_P^2\,H_0^2 \,, \qquad {B} =
\frac{\nu\,M_P^2}{24\pi}\nonumber\,. \eeq { In writting the
cosmological term as in equation (\ref{A and B}), we intend to
obtain a covariant expression which reproduces the background
relations. Although this expression may not be unique (in the sense
that additional terms vanishing in the background could nevertheless
contribute to the perturbations), any other covariant form
reproducing that behavior would be more involved. Therefore, we will
adopt the simplest possibility (\ref{A and B}) as our ansatz for the
calculation.} Any variable cosmological term that varies as the
square of the Hubble function could be written in the same manner.
However, the concrete way such term appears in the background, and
the relation of the factors ${A}$ and ${B}$ to the quantum field
mass through the parameter $\nu$ -- see Eq.\,(\ref{nu})-- is
specific of the present model.
\par
Consider next simultaneous perturbations of the two densities and the
metric
\beq
\rho_m \to \rho_m \,(1 + \de_m)\,, \qquad\qquad
\rho_\La \to \rho_\La \,(1 +  \de_\La)\,, \qquad\qquad
g_{\mu\nu} \to g^\prime_{\mu\nu} = g_{\mu\nu} + h_{\mu\nu}\,.
\label{perturbations}
\eeq
The background metric \
$g_{\mu\nu}=\mbox{diag}\, \big\{1,\,-a^2(t)\,\de_{ij}\big\}$
corresponds to the solution described in the prrevious section.
We assume the
synchronous coordinate condition $h_{0\mu}=0$ and obtain the
variation of the $R_0^0$ in the form
\beq
\de R^0_0 = \frac12\,\frac{\pa^2}{\pa t^2}\,
\left(\frac{h_{ii}}{a^2}\right) +
H \,\frac{\pa}{\pa t} \left(\frac{h_{ii}}{a^2}\right)\,.
\label{0 Ricci perturbations}
\eeq
It proves useful to introduce the
notation
\begin{equation}
\hh \equiv \frac{\pa}{\pa t} \left(\frac{h_{ii}}{a^2}\right)\,.
\label{hdef}
\end{equation}
This variable satisfies the equation
\beq \dot {\hh} + 2H\,\hh =
8\pi G\,\big(\de\rho_m-2\de\rho_\La \big)\,. \label{first}
\eeq
Perturbing the covariant conservation law, $\na_\mu T^\mu_\nu
= 0$, we arrive at the following equation for the $\nu=0$
component:
\beq
 \de {\dot \rho}_m + \big(\th -
\frac{\hh}{2}\big)\,\rho_m + 3H \,\de\rho_m\,=\,-\,\de {\dot
\rho}_\La\,, \label{zero .} \eeq and, for the $\nu=i$ components,
at the equation \beq {\dot \rho}_m \th + 5H\rho_m\th + \rho_m
{\dot \th} \,=\,- \frac{k^2}{a^2}\,\de \rho_\La \,. \label{k .}
\eeq Here we have used the constraint $\de U^0=0$, and also the
notation $\,\nabla_i (\de U^i)\equiv \th\,$ for the covariant
derivative of the perturbed $3$-velocity. We should not confuse
the wave number  $k$ in Eq.\,(\ref{k .}) and below with the
spatial curvature parameter introduced before, by the context it
should be obvious. Indeed, it is understood that we have written
all the previous perturbation equations in Fourier space, namely
using the standard Fourier representation for all quantities:

\beq
f({\vec x},t)\,=\,\int \frac{d^3k}{(2\pi)^3}\,\,
\tilde{f}({\vec k},t)\,e^{i{\vec k}\cdot{\vec x}}\,, \qquad
k=\left|{\vec k}\right|. \label{Furier} \eeq
 Perturbing the Eq.\,(\ref{A and B}) and
using the relation (\ref{H}) one finds \beq \rho_\La \cdot \de_\La
\,=\,\de\rho_\La \,=\, 2{B}\,\Big(\th-\frac{\hh}{2}\Big)(\na_\mu
U^\mu) = 6H{B}\,\Big(\th-\frac{\hh}{2}\Big)\,. \label{last} \eeq

It proves useful to rewrite the perturbations in terms of another
set of variables. Let us introduce the new variable \ $v =
{\th\rho_m}/{\rho_t}=\th\,f_1$  and trade the time derivatives for
the ones in the redshift parameter  $z$, using the formula
$d/dt=-(1+z)\,H\,d/dz$. After a straightforward calculation the
perturbation equations can be cast into the following form:
\beq
v^\prime \,+ \,\frac{\left(3f_1 - 5\right)}{1+z}\,v
\,=\,\frac{k^2\,(1+z)\,f_2}{H}\,\de_\La \,, \label{zero ...} \eeq
\beq \de^\prime_m + \left(\frac{f_1^\prime}{f_1} -
\frac{3f_2}{1+z}\right)\,\de_m -
\frac{1}{(1+z)H}\,\left(\frac{v}{f_1} - \frac{\hh}{2}\right)
\,=-\,\frac{1}{f_1}\,\left(f_2\,\de_\La\right)^\prime -
\frac{3\,f_2}{1+z}\,\de_\La\,, \label{k ...} \eeq
 \beq
\hh^\prime - \frac{2\hh}{1+z}
\,=\,-\,\frac{2\nu}{(1+z)\,\varrho}\, \left(f_1\,\de_m -
2\,f_2\,\de_\La\right)\,, \label{h ...} \eeq \beq \de_\La
\,=\,\frac{\varrho}{f_2}\, \left(\frac{v}{f_1} -
\frac{\hh}{2}\right)\,. \label{2} \eeq In the last two equations
we have introduced the function \beq \varrho =
\frac{2\,\nu\,H(z)}{3H^2(z)-3H_0^2\Omega_k^0(1 + z)^2} \,.
\label{g function} \eeq The equation (\ref{2}) is not dynamical,
rather it represents a constraint which can be replaced in the
other equations (\ref{zero ...})-(\ref{h ...}). Using notations
(\ref{ratios}) and (\ref{g function}) we eventually arrive at:
\beq v'\,+\,\frac{(3f_1-5)}{1 + z} \,v \,-\,
\frac{k^2\,\varrho\,(1+z)}{H\,f_1}\,v \,=\, -
\,\frac{k^2\,\varrho\,(1+z)}{2\,H}\,\hh \,; \label{e1} \eeq \beq
\hh' \,+\, \frac{2(\nu-1)}{1 + z}\,\hh \,=\,\frac{2\,\nu}{(1 +
z)}\, \Big(\,\frac{2\,v}{f_1}\, -
\,\frac{f_1}{\varrho}\,\de_m\,\Big) \,; \label{e2} \eeq \beq
\delta'_m + \Big(\frac{f_1^\prime}{f_1} -
\frac{3f_2}{1+z}\Big)\,\delta_m \,=\, \frac{1}{f_1}\,
\Big(\frac{\varrho\,\hh}{2}-\frac{\varrho\,v}{f_1}\Big)^\prime
\,+\,\frac{1}{1+z}\,\Big(\,3\varrho - \frac{1}{H}\,\Big)\,
\Big(\,\frac{\hh}{2}-\frac{v}{f_1}\,\Big)\,. \label{e3} \eeq

These equations constitute the complete set of coupled Fourier
modes for the velocity, density and metric perturbations in the
presence of a running cosmological term. One can check that for
vanishing $\nu$ they reproduce the situation in the $\Lambda$CDM
model\,\cite{peebles,LL,weinberg}, as expected from the fact that
$\nu=0$ corresponds to having a strictly constant $\CC$ and
conserved matter density $\rho_m$. The previous set of equations
are in the final form to be used in our numerical analysis of the
density perturbations for arbitrary values of $\nu$ and their
impact on the large scale structure formation in the present
Universe\,\footnote{In Ref.\,\cite{FabrisSpindel} perturbations
were considered in the very early Universe for an inflationary
model with exponentially decaying vacuum energy under the
assumption of a thermal spectrum and with no obvious relation to
QFT and RG.}.

\vspace{0.3cm}

%%%%%%%%%%%%%%%%%%%%%%%%%%%%%%%%%%%%%%%%%%%%%%%%%%%%%%%%%%%%%%%%%%
\section{Power-spectrum analysis and comparison
with the galaxy redshift survey}
\label{sect:RGmodel3}

\subsection{Power spectrum and transfer function} \label{sect:powertransfer}

In the previous section we have obtained the coupled set of
differential equations (\ref{e1}), (\ref{e2}) and (\ref{e3})
describing the dynamics of the density and  metric perturbations.
In order to perform the numerical analysis of these equations, we
have to fix the initial conditions and also define the limits for
the variation of the redshift parameter $z$ and for the
perturbations wave number $k=\left|{\vec k}\right|$. The
analysis performed here applies after the radiation dominated era,
in fact $z$ must vary from about $z = 1100$ (the recombination era)
to $z = 0$ (today). For the sake of the numerical analysis we measure
$k$ in the units of $h\,$Mpc$^{-1}$,  where $h$ is the
reduced Hubble constant. In these units let us notice that we must
 consider $k <0.15\,h\,$Mpc$^{-1}$
since the observational data concerning the linear regime are in
this range\,\cite{cole}. Expressing the scales in terms of the
Hubble radius $H_0^{-1}=\simeq 3000\,h\,$Mpc implies that we must
consider the values of $k$ between 0 and $450\,{H_0}$.

The expression for the baryon matter power spectrum is given in
many places of the literature\,\cite{peebles,LL}. We use the
concrete form given in \cite{Martin}. At z=0 we
have\,\footnote{In general the power spectrum depends on a
parameter $n$ labeling the possibility of a non-vacuum initial
state of the perturbations of quantum origin\,\cite{Martin}. We
take of course $n=0$, corresponding to the power spectrum of the
vacuum.}
\begin{equation}
\label{Powers} P(k)
\,\equiv\Big|\frac{\delta\rho_m(k)}{\rho_m(k)}\Big|^2 \,=\,
|\delta_m(k)|^2 \,=\,
A\,\,k\,\,T^2(k)\,\frac{g^2(\Om_T^0)}{g^2(\Om^0_{M})}\,,
\end{equation}
where $\Om_T^0\equiv\Om_{M}^0+\Om_\La^0$, with
$\Om_{M}^0=\Om_{DM}^0+\Om_B^0$ as defined before. (Notice that for flat
geometry $\Om_T^0=1$.) Here $A$ is a normalization coefficient
(see below), $T(k)$ is the transfer function and $g(\Omega)$ is
the growth function -- see below for details on these functions.
We consider a scale invariant (Harrison-Zeldovich) primordial
scenario as revealed by the assumed linear dependence in $k$ of
(\ref{Powers}). The transfer function takes into account the
physical processes occurring around the recombination era. The
use of the transfer function allows to perform the numerical
analysis and the comparison with the recent data from the
2dFGRS\,\cite{cole} deep in the matter dominated era, for example
in the range of $z$ between $z_0=500$ and $z_f=0$. The choice of
$z_0$ is quite arbitrary, the only requirement is that the system
could evaluate sufficiently deep into the matter dominated phase.
Essentially the results do not depend on the precise value of
$z_0$.

In order to fix properly the initial conditions, we use the
transfer function presented in references \cite{Bardeen,Martin}.
This transfer function assumes a scale invariant primordial
spectrum, and determines the spectrum today considering the
Universe composed of dark matter and a cosmological constant that
generates the actual accelerated expansion phase. With the aid of
this transfer function we can fix the initial conditions at a
redshift after the recombination. Of course, our model contains
the traditional cosmological constant as a particular case when
our relevant RG parameter $\nu = 0$, see Eq.\,(\ref{A and B}).

Let us now specify the transfer function. To this end we use
the following notations \cite{Martin}:
\beq
\Ga\,=\,\Om_{M}^0\,h\,
e^{-\,\Om_B^0\,-\,\left(\Om_B^0/\Om_{M}^0\right) } \,,\quad
q=q(k)=\frac{k}{(h \Ga)\,{\rm Mpc}^{-1}}\,.
\eeq
In these expressions, $\Ga$ is
Sugiyama's shape parameter\,\cite{Sugiyama95}. Given an initial spectrum (e.g.,
Harrison-Zeldovich), the power spectrum today can be obtained by
integrating the coupled system of equations for the evolution of
the Universe and the Boltzmann equations. The transfer function
for the baryonic power spectrum can be approximated by the
following numerical fit (BBKS transfer function) \cite{Bardeen}:
 \beq \label{jtf}
T(k)=\frac{\ln (1+2.34 q)}{2.34 \,q(k)}\, \left[1+3.89 q +
\left(16.1 q\right)^2 + \left(5.46 q\right)^3 + \left(6.71
q\right)^4\right]^{-1/4}\,. \eeq
Furthermore, the growth function
appearing in (\ref{Powers}) reads\,\cite{CPT92}
\beq\label{CPT}
g(\Om)=\frac{5\Om}{2}\,
\left[\Om^{4/7} - \Om_{\La} + \Big(1+\frac{\Om}{2}\Big)
\Big(1+\frac{\Om_\La}{70}\Big)\right]^{-1}\,.
\eeq
This function
takes into account the effect of the cosmological constant in the structure
formation (a suppressing effect when the CC is positive) .
Notice that both the transfer function and the
growth function are involved in the computation of the matter
power spectrum in Eq.\,(\ref{Powers}).
%%%%%%%%%%%%%%%%%%%%%%%%%%%%%%%%%%%%%%%%%%%%%%%%%%%%%%%%%%%%%%%%%
Finally, the normalization coefficient $A$ in (4.1) can be fixed
using the COBE measurements of the spectrum of CMB anisotropies.
This coefficient is connected to the quadrupole momentum $Q_{rms}$
of the CMB anisotropy spectrum by the relation \cite{Martin} \beq A
\,=\, (2l_H)^4\, \frac{6\pi^2}{5}\,\frac{Q^2_{rms}}{T_0^2} \,,
\label{AA} \eeq where $l_H\equiv H_0^{-1}\simeq 3000h^{-1}\,$Mpc is
the present Hubble radius and $T_0 = 2.725\pm0.001\,^{o}K$ is the
present CMB temperature. The quadrupole anisotropy will be taken as
\begin{equation}
Q_{rms} = 18 \mu\,K \,.
\end{equation}
This value is obtained from COBE normalization
and is consistent with the more recent results
of the WMAP measurements \cite{hinshaw}
using the prior of a scale invariant spectrum.
% \footnote{Let us notice that it is consistent with the
% more recent results
% of the WMAP measurements \cite{spergel,hinshaw} \
% $Q_{rms} = 15.3^{+3.8}_{-2.8} \mu K  %%\label{Q}
% $.}
Taking all these estimates into account, we can fix
 \beq A = 6.8\times 10^{5}\,
{\rm Mpc}^4\,, \label{A}
\eeq
which corresponds to the
initial vacuum state for the density perturbations used in
\cite{Martin}.
%% As we can see,
%% the fluctuations of the CMB temperature enable one to fix
%% entirely the normalization of the mass spectrum for the
%% given choice of the transfer function.

There is evidently nothing sacred about this particular choice
of the transfer function. Many others have been proposed in the
literature. For example, the following useful simplified transfer function
has been proposed by Peebles in Ref. \cite{peebles}:
\beq
T(k) = \Big[\,1\,+\,\frac{8\,k}{\Omega_M^0\,h^2}
\,+\,\frac{4.7\,k^2}{(\Omega_M^0\,h^2)^2}
\,\Big]^{-1}\,.
\label{spectrum}
\eeq
In this function it is understood (as in the BBKS one (\ref{jtf})) that $k$ is given in
$h\,$Mpc$^{-1}$ units (as mentioned above).
Strictly speaking the normalization used in (\ref{spectrum}) should be
re-computed for the present model using the CMB anisotropy spectrum,
but this calculation is very involved and lies beyond the scope of the present
work. Let us notice that the remaining arbitrariness in
the normalization of the power spectrum may not affect the
qualitative results of our
investigation. In particular, we have checked that the results
obtained with Peeble's function (\ref{spectrum}) are very
close to those obtained with the BBKS function (\ref{jtf}).
However, the last one is more detailed, and we can control better
the contribution of each component of the matter content for the
final power spectrum. For this reason we present here only the results
for the transfer function (\ref{jtf}).

\subsection{Initial conditions. Normalization with respect to to the
 $\Lambda$CDM model} \label{sect:initialcond}

Let us now explain in some more detail how do we fix the initial
conditions for our specific problem. The strategy is the
following. We first consider the standard $\Lambda$CDM
cosmological constant case as a reference to normalize our
calculation. In this way the baryonic spectrum today (that is, for
$z = 0$) can be described by e.g. the transfer function
(\ref{jtf}). This allows to fix the conditions for the density
contrast $\delta_m$ at $z=0$ for the $\Lambda$CDM case. Moreover,
for the initial condition on the metric function, $\hat{h}$, we
suppose that it is identical to the density contrast, because for
the $\Lambda$CDM model this is so (up to a small factor). Finally,
for the velocity perturbations, $\,\th\,\equiv\nabla_i (\de U^i)$,
we suppose that they are zero today, since they are decreasing
functions and they are precisely zero for the strict $\Lambda$CDM
model. With this initial $\Lambda$CDM normalization, we may
proceed to find the spectrum for the RG model in two steps.
First, we integrate back the perturbed equations in the
$\Lambda$CDM case from the present time ($z=0$) until some point
\ $z = z_0$, where \ $z_0 \gg 1$. As we said, it is not very
important the particular value of $z_0$ (say $500$, $400$ etc)
provided it lies well after the recombination era, i.e.
$z_0<1100$. In this way we determine the values of the three
fluctuations $(\delta_m,\hat{h},v)$ at $z=z_0$. Second, we can
then use these values as initial conditions for the RG case.
Indeed, since for \ $z_0 \gg 1$ the cosmological constant does
not play an important role, we can use the initial conditions
defined in this manner to solve the perturbed equations
(\ref{e1})-(\ref{e1}) in our RG model for arbitrary $\nu$. In
particular, this provides the output for the baryonic density
spectrum today, $\delta_m(k)$, for $\nu\neq 0$. From here the
corresponding matter power spectrum (\ref{Powers}) can be
immediately evaluated. Notice that the comparison with the
cosmological constant case makes sense, since all the difference
between the running cosmological constant and the particular case
of having a strictly constant $\Lambda$ (corresponding to the
standard $\Lambda$CDM model) lies in the evolution of the
Universe for a relatively small value of $z$. In fact, the
contribution of the dark energy component begins to become
relevant for small redshifts, say for $z < 10$ or even less, and
the running is sizeable if $|\nu|$ is not smaller than $10^{-3}$.
To summarize, we use the well known semi-analytic expressions for
the standard cosmological constant case (which can be reproduced
from the particular case $\nu=0$ of our general RG framework) to
settle the initial conditions for our running cosmological
constant scenario ($\nu\neq 0$).

The final baryonic spectrum depends essentially on the following
three parameters: the relative amounts (with respect to the
critical density) of baryonic matter, dark matter and dark energy
today, i.e. \ $\Omega_B^0$, \ $\Omega_{DM}^0$ and
$\Omega_{\Lambda}^0$ respectively. The curvature today is given
by $\Omega_{k}^0 = 1 - \Omega_{M}^0 - \Omega_{\Lambda}^0$, where
of course $\Omega_{M}^0=\Omega_{B}^0 + \Omega_{DM}^0$. Our aim is
to compare the theoretical spectrum with the LSS data from the 2dF
Galaxy Redshift Survey\,\cite{cole}, with the error bars
evaluated at the $1\sigma$ level. To start with we consider the
standard $\Lambda$CDM model case (therefore we set the relevant RG
parameter $\nu=0$ in our perturbations equations) and produce some
numerical examples for the three spatial curvatures. The corresponding
results for the open ($\Omega_{k}^0 = 0.6$),  flat ($\Omega_{k}^0
= 0$) and closed ($\Omega_{k}^0 = - 0.5$) cases, all of them with
\ $\Omega_{DM}^0 = 0.21$ \ and \
$\Omega_{B}^0 = 0.04$ \ (as suggested by primordial
nucleosynthesis), using the transfer function (\ref{jtf}), are
depicted in Fig.\,\ref{fig1}a,b,c respectively. The presented plots differ
essentially by the amount of dark energy and it can be easily
seen that the best fit corresponds to the flat geometry case (the plot
in Fig. \,\ref{fig1}b), with \
$\Omega_{\Lambda}^0=0.75$.  This was the expected result and
therefore this preliminary exercise serves us as a good normalization of
our computation.

%%%%%%%%%%%%%%%%%%%%%%%%%%%%%%%%%%%%%%%%%%%%%%%%%%%%%%%%%%
\FIGURE[t]{
\mbox{\resizebox*{0.45\textwidth}{!}{\includegraphics{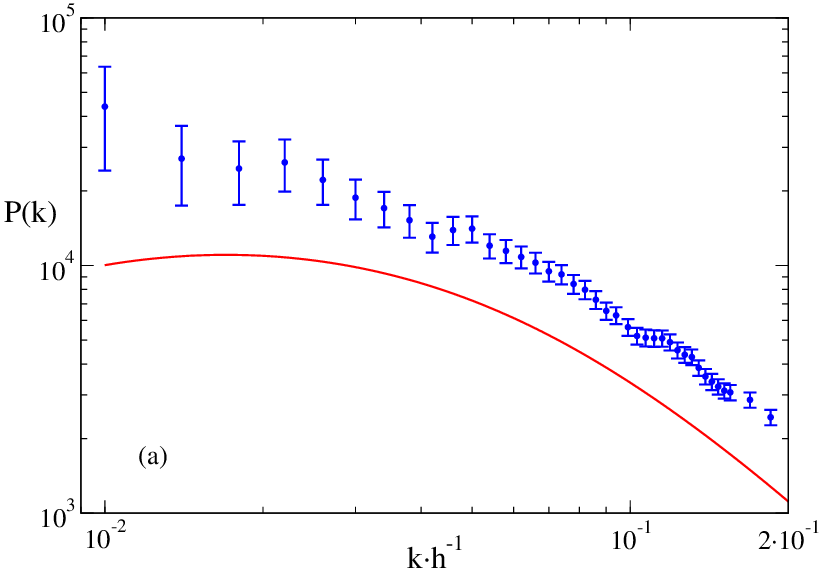}}
\ \ \ \
   \resizebox*{0.45\textwidth}{!}{\includegraphics{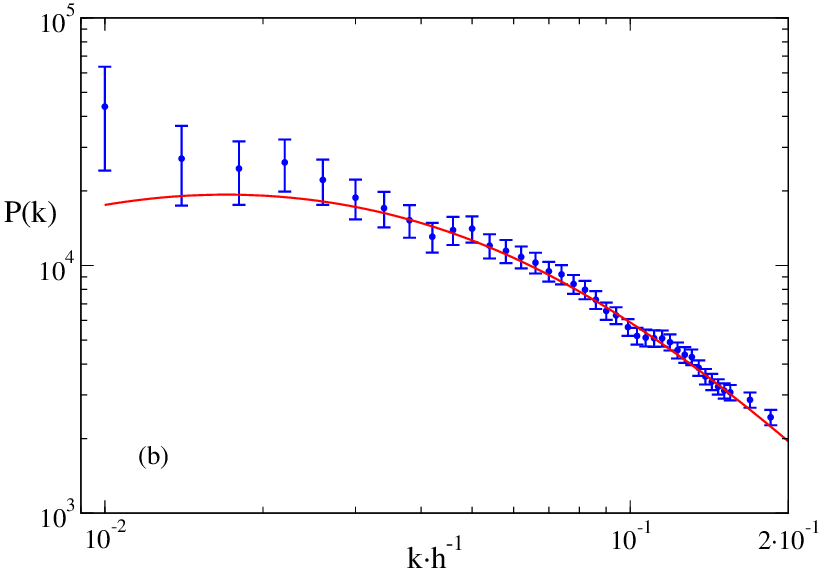}}}
\\
   \\
   \\
   \mbox{\resizebox*{0.45\textwidth}{!}{\includegraphics{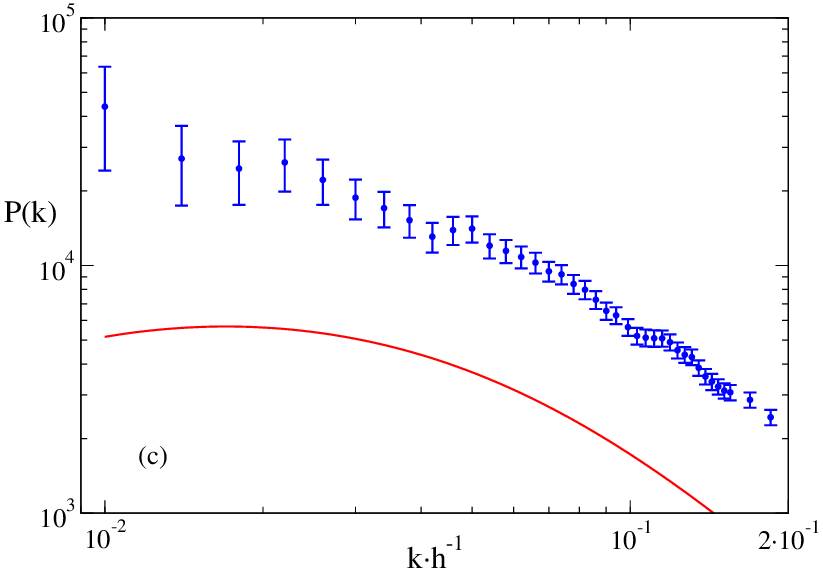}}}
\caption{ Baryonic power spectrum for the $\Lambda$CDM model, for
fixed $\Omega_{B}^{0}=0.04$, $\Omega_{DM}^{0}=0.21$ and for (a)
$\Omega_{\Lambda}^{0}=0.15$ (open Universe), (b)
$\Omega_{\Lambda}^{0}=0.75$ (flat Universe) (b) and (c)
$\Omega_{\Lambda}^{0}=1.25$ (closed Universe), together with the
LSS data from the 2dfFGRS\,\cite{cole}. The ordinate axis
represents  $P(k)=|\delta_m(k)|^2$ at $z=0$ as given by
(\ref{Powers}) with the transfer function (\ref{jtf}) and growth
function (\ref{CPT}), while in the abscissa we have the wave
number $k$ given in $h\,$Mpc$^{-1}$ units. } \label{fig1}}
%%%%%%%%%%%%%%%%%%%%%%%%%%%%%%%%%%%%%%%%%%%%%%%%%%%%%%%%%%%%%%

In Fig.\,\ref{fig2} we fix the values  $\Omega_{B}^0 = 0.04$ and
$\Omega_{\Lambda}^0 = 0.75$  and vary the dark matter content
$\Omega_{DM}^0$ in each plot.  Specifically, we plot the three
cases $\Omega_{DM}^0 = 0.01,\, 0.21$ \ and \ $0.41$, representing
an open, flat and closed Universe, respectively. We note that the
pressureless components of the matter content (dark matter and
baryons) tend to increase the power spectrum.  One can see
(for the given values of $\Omega_{B}^0$ and $\Omega_{\Lambda}^0$)
that the best fit occurs for the expected DM content, namely around
$\Omega_{DM}^0 =0.21$ (cf. Fig.\,\ref{fig2}b). This provides
further confidence in the normalization of our computation with
respect to the standard $\Lambda$CDM case.

%%%%%%%%%%%%%%%%%%%%%%%%%%%%%%%%%%%%%%%%%%%%%%%%%%%%%%%%%%%%%%%%%%%
\FIGURE[t]{
\mbox{\resizebox*{0.45\textwidth}{!}{\includegraphics{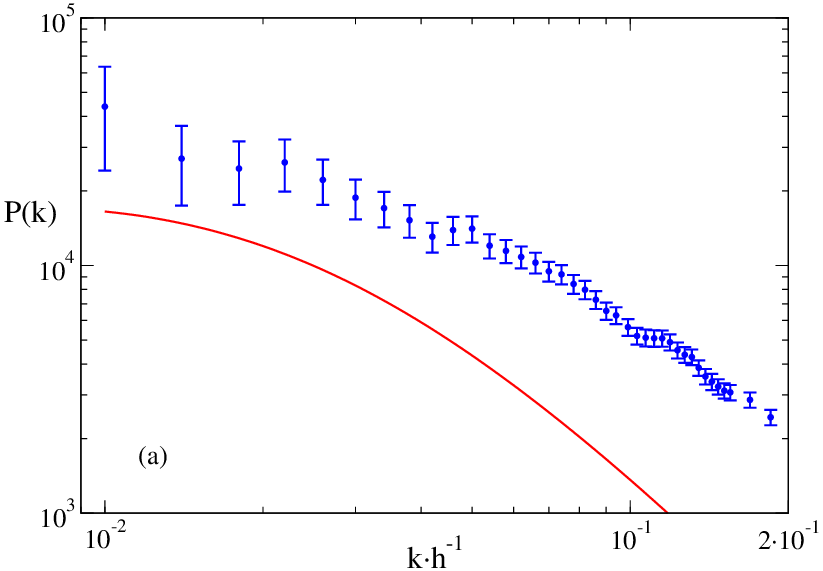}}\
\ \ \
   \resizebox*{0.45\textwidth}{!}{\includegraphics{log1b.eps}}}\\

   \\
   \\
   \mbox{\resizebox*{0.45\textwidth}{!}{\includegraphics{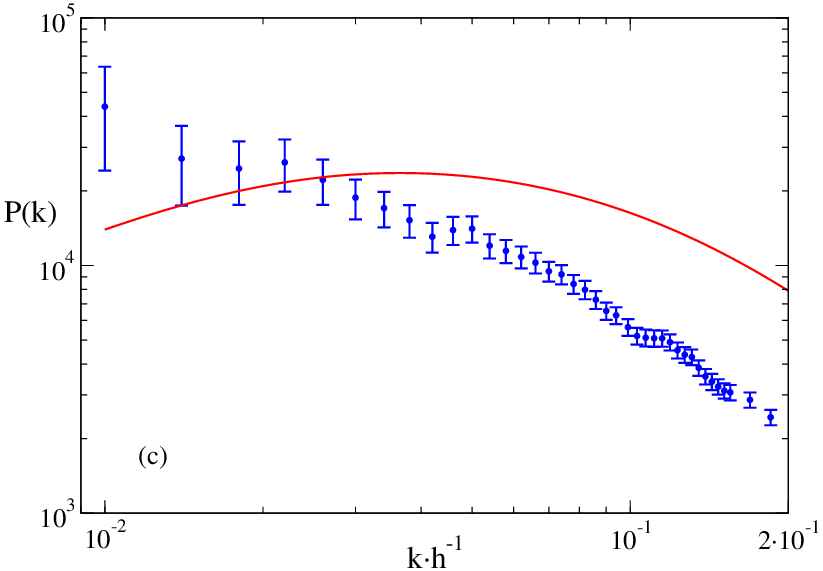}}}
\caption{ As in Fig\,\protect\ref{fig1} ($\Lambda$CDM model) for fixed
$\Omega_{B}^0 = 0.04$,
$\Omega_{\Lambda}^0 = 0.75$ and for (a) $\Omega_{DM}^0 = 0.11$ (open Universe),
(b) $\Omega_{DM}^0=0.21$ (flat Universe) and (c) $\Omega_{DM}^0=0.41$ (closed Universe),
together with the LSS data from the 2dfFGRS\,\cite{cole}. } \label{fig2}}
%%%%%%%%%%%%%%%%%%%%%%%%%%%%%%%%%%%%%%%%%%%%%%%%%%%%%%%%%%%%%%%%%%%%%%%%%%

\subsection{Power spectrum for the running $\Lambda$ case} \label{sect:powerspectrum}

Now we can start to deal with our main task, that is, to explore
the influence of the RG parameter $\nu$ on the processed power
spectrum. It is apparent that the larger is the value of $\nu$,
the more important is the contribution of the function
$\varrho(z)$, Eq.(\ref{g function}), in the solution of our set
of differential equations. This function appears as a factor of
the Laplacian acting on the perturbed quantities, see
Eq.\,(\ref{k .}). Hence, large $\varrho(z)$ entails a more
intensive damping of the perturbations contributing to depress
the power spectrum. This feature can be seen at work in the plots
of Fig.\,\ref{fig3}a,b,c,d. These plots illustrate the power
spectrum for running $\Lambda$ in the flat case and for
non-vanishing and positive $\nu$. They correspond to
$\Omega_{DM}^0 = 0.21$, $\Omega_{B}^0 = 0.04$ and $\Omega_\La^0 =
0.75$, for \ $\nu = 10^{-8}, 10^{-6}$, $10^{-4}$ \ and \
$10^{-3}$ \ respectively. We can easily grasp the impact of the
$\nu$ parameter, namely we see quite patently that the larger is
$\nu>0$ the more suppressed is the power, mainly at small scales
(i.e. at large $k$). For values of $\nu>0$ up to $\nu = 10^{-6}$
the agreement with the LSS data is almost perfect; for $\nu =
10^{-5}$ (not shown in the figures) the theoretical curve starts
to depart maily in the high $k$ region, i.e. at small
distances. Above \ $\nu = 10^{-4}$ the spectrum presents a strong
deviation (depletion of the power) with respect to the
observational data, as can be seen in Fig.\,\ref{fig3}d
corresponding to $\nu = 10^{-3}$. This is
a quite robust result, and we can state that values of $\nu$
above $10^{-4}$ are definitely ruled out by the perturbations
analysis.

The interpretation of this result in terms of particle
physics can be obtained from Eq.\,(\ref{nu}), in which we recall
that $M$ stands for the effective mass of the heavy particles
contributing to the running of the cosmological term. Therefore,
in the framework of the given model for the running CC -- that is,
assuming the possibility of an energy exchange between vacuum and
matter sector via the conservation equation (\ref{conservation})
-- the exclusion of values $\nu> 10^{-4}$ means that we can rule
out the existence of particles with masses of the Planck mass
order, unless their contributions to the CC cancel quite
accurately due to some strong symmetry (e.g. supersymmetry).

%%%%%%%%%%%%%%%%%%%%%%%%%%%%%%%%%%%%%%%%%%%%%%%%%%%%%%%%%%%%%%%%%%%%%%%
\FIGURE[t]{
\mbox{\resizebox*{0.45\textwidth}{!}{\includegraphics{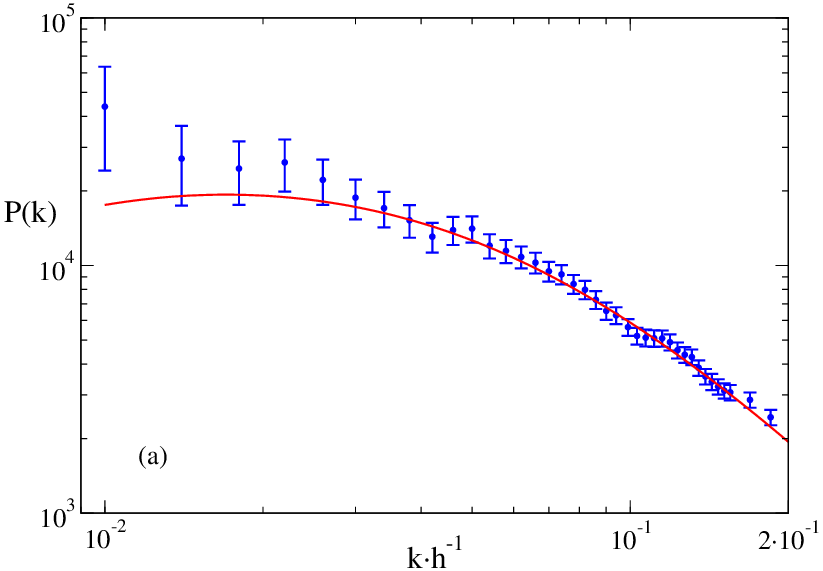}}\
\ \ \
   \resizebox*{0.45\textwidth}{!}{\includegraphics{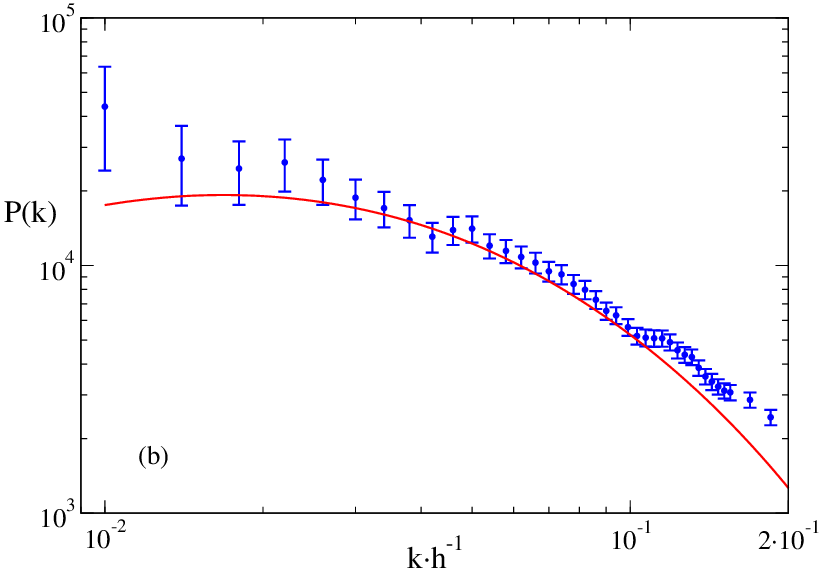}}}\\

   \\
   \mbox{\resizebox*{0.45\textwidth}{!}{\includegraphics{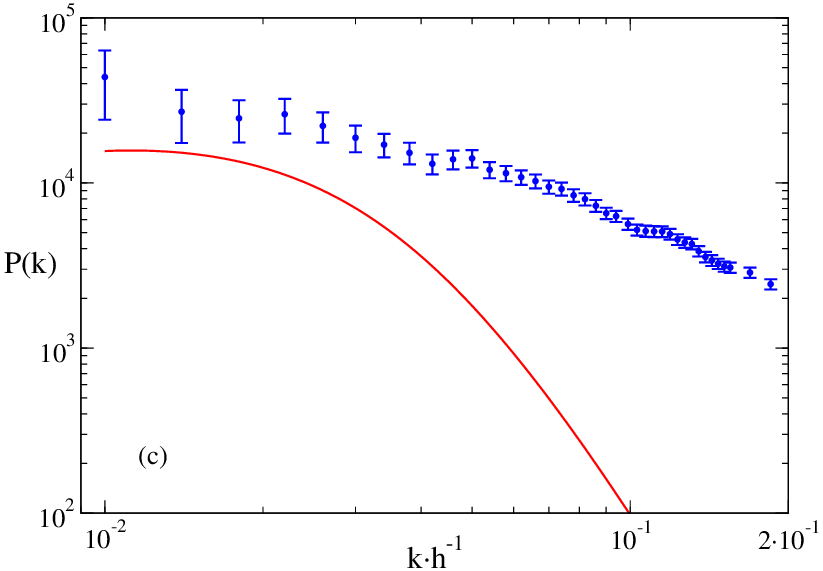}}\
\ \ \
   \resizebox*{0.45\textwidth}{!}{\includegraphics{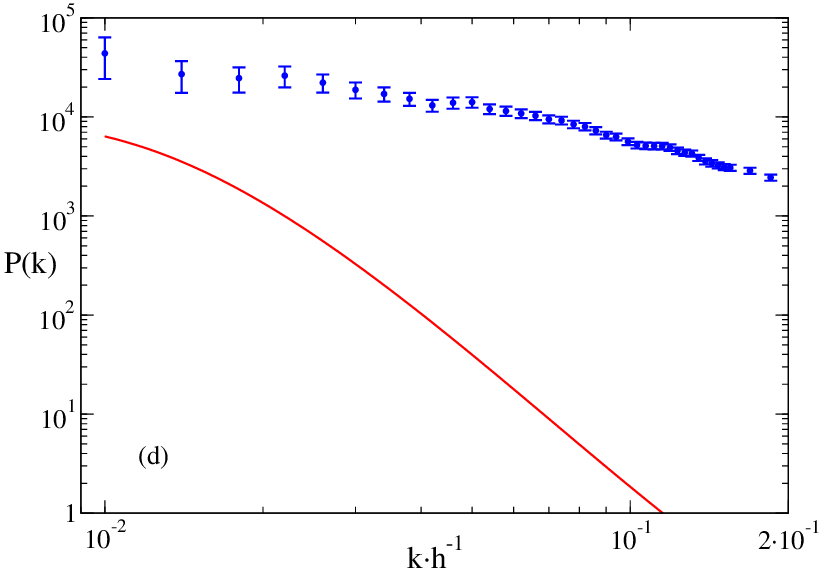}}}
\caption{As in Fig\,\protect\ref{fig1}, but for the \
$\nu$-dependent power spectrum of the running cosmological
constant model, together with the LSS data from the
2dfFGRS\,\cite{cole}. The ordinate axis represents
$P(k)=|\delta_m(k)|^2$ where $\delta_m(k)$ is the solution of
(\ref{e1})-(\ref{e3}) at $z=0$. In all cases we have fixed
$\Omega_{B}^0 = 0.04$, $\Omega_{DM}^0 = 0.21$ and
$\Omega_{\Lambda}^0 = 0.75$ with the following (positive) values
of $\nu$: (a) $10^{-8}$, (b) $10^{-6}$, (c) $10^{-4}$ and (d)
$10^{-3}$.
  } \label{fig3}}
%%%%%%%%%%%%%%%%%%%%%%%%%%%%%%%%%%%%%%%%%%%%%%%%%%%%%%%%%%%%%%%%%%%%%%%

On the face of the previous result, the next relevant question is:
which is then the minimum upper bound on $\nu$ compatible with the
observational data?  It turns out that for values of $\nu>0$
between $10^{-6}$ and $10^{-4}$ (and above all between $10^{-6}$
and $10^{-5}$) it is not possible to state, at this stage, if
they are definitely ruled out or not. In principle, from the
inspection of Fig.\,\ref{fig3} we could strongly argue in favor
of the safer range $0\leqslant\nu \leqslant 10^{-6}$. However, we
must take into account the potential effect of other components.
For instance, if we increase the amount of dark matter (which is
not excluded), the values of \ $\nu > 10^{-6}$ \ may lead to a
better fit.
%
% This is exemplified
% by figures
% $3$, where the quantity of dark energy, as well as of baryons, is
% fixed, while the amount of dark matter is increased. Better fits
% are obtained for higher values of $\nu$. In this case, a closed
% Universe is preferred.
%
On the other hand, for negative values of $\nu$ the bound becomes
essentially stronger, i.e. $\nu<0$ values  produce larger
deviations beyond a critical range of $|\nu|$. In practice this
means that for $\nu$ outside the interval $-10^{-6}\leqslant
\nu\leqslant 0$ the spectrum deviates from observations more
strongly than for positive values of $\nu$ outside the range
$0\leqslant \nu\leqslant 10^{-6}$. This is because the parameter
$\nu$ \ is multiplied, in the Eq.\,(\ref{e1}), by the
$k^2$-proportional terms which stem from the pressure gradient in
the perturbed equations. When the parameter $\nu$ becomes
negative, the sign of the corresponding coefficient changes. This
reflects in the evaluated power spectrum in the following way:
while for a positive $\nu$ the spectrum is depressed in the small
scale regime, for negative $\nu$ the spectrum is enhanced in that
regime, and for a value of $\nu$ well below $-10^{-6}$ the
spectrum becomes rapidly divergent there. This behaviour is
exemplified in Fig.\,\ref{fig4}a,b,c for the cases \ $\nu = -
10^{-8}$, $- 10^{-6}$ and $\ -10^{-4}$ respectively. In
Fig.\,\ref{fig5} we further elaborate on the $\nu<0$ features
studied in the previous figure, namely we superimpose the
detailed evaluation of the spectrum for the five cases $\nu = -
10^{-8}, -10^{-7},-10^{-6}, -10^{-5}$ and $- 10^{-3}$. In this
way we can better appreciate the aforementioned divergence at
small scales, which becomes more and more evident the larger is
$|\nu|$. In fact, in this case we see that for $|\nu|>10^{-5}$ the
deviation is far more pronounced than for $\nu>0$ and this
demonstrates that the departure of the predicted spectrum from the
experimental values in the range $\ -10^{-4}\leqslant\nu\leqslant
-10^{-5}$ is much more violent than in the corresponding positive
range $\ 10^{-5}\leqslant\nu\leqslant 10^{-4}$.

The change of behaviour in passing from positive to negative
values of $\nu$ could be expected. It corresponds to the change of
an oscillatory behaviour into an exponential one or, in other
words, to the transition from a harmonic oscillator regime to an
anti-harmonic oscillator regime. We can also understand why the
spectrum is not essentially affected at large scales by the change
of sign in $\nu$, since at large scales the pressure gradient is
negligible. For negative $\nu$, the variation of the amount of
dark matter and dark energy affect the final spectrum in the same
way as for a positive $\nu$ case.

To summarize, the approximate range of values of the RG parameter
which are amply allowed by the analysis of the matter power
spectrum of perturbations (i.e. what we may call the ``safest
range'' of allowed values of $\nu$) is the following:
\begin{equation}\label{nurange}
-10^{-6}\leqslant \nu\leqslant 10^{-6}\,.
\end{equation}
There are nevertheless some $\nu$ intervals outside the safest
region which can still be considered allowed (or at least not
ruled out) by the density perturbations analysis. For instance,
positive values of $\nu$ in the range $10^{-6}-10^{-4}$ (mainly
those between $10^{-6}$ and $10^{-5}$) cannot be definitely ruled
out; we can only say that the corresponding spectrum starts to
show some deviation from the LSS data becoming more and more acute
the closer we get to $10^{-4}$. For $\nu<0$ the exclusion effect
is stronger, and indeed we can safely rule out the range
$\nu<-10^{-5}$ for which the predicted spectrum becomes rapidly
divergent at low scales. Furthermore, for $|\nu|>10^{-4}$ the
deviation is markedly patent for any sign of $\nu$ and we can
assert that this range is manifestly excluded.

Finally, we may ask ourselves how to better improve the aforesaid
limits on the RG parameter $\nu$. In our opinion this can only be
accomplished by the simultaneous crossing of the LSS data with
further observational data, most significantly with the wealth of
CMB anisotropy data. This should constraint much better the value of
$\nu$ beyond the LSS limits that we have been able to obtain. In
particular, it would probably help to decide what is the situation
with the unsettled positive range $\nu=10^{-6}-10^{-4}$ where the
LSS data cannot give a last word. In principle, the CMB spectrum
predicts an almost spatially flat Universe. But this result is model
dependent and another full fledged direct evaluation of the
cosmological parameters using the running cosmological constant
model must be made. We plan to perform this analysis in the future.

%%
%% However, for the values \ $\nu < - 10^{-6}$, the divergence
%% can not be erased by varying the fractions of
%% matter components.

%%%%%%%%%%%%%%%%%%%%%%%%%%%%%%%%%%%%%%%%%%%%%%%%%%%%%%%%%%%%%%%%%%%%%%%%%%
\FIGURE[t]{
\mbox{\resizebox*{0.45\textwidth}{!}{\includegraphics{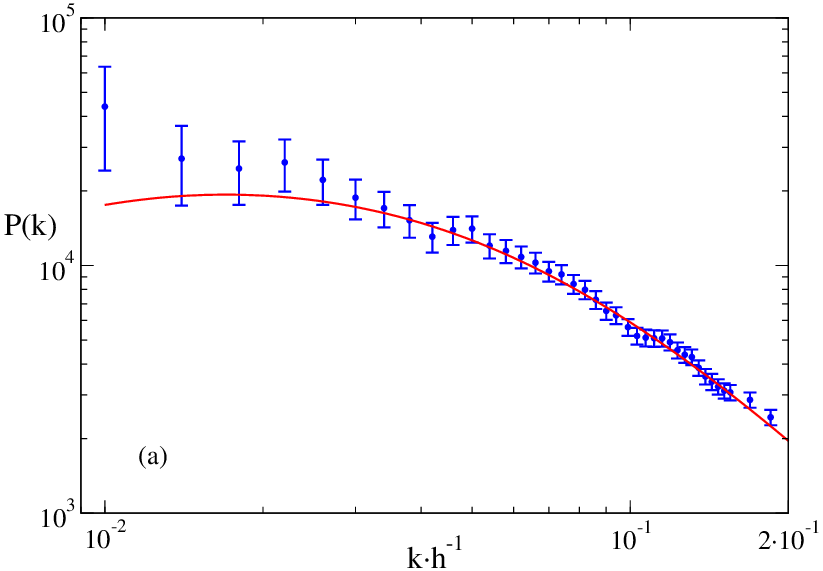}}\
\ \ \
   \resizebox*{0.45\textwidth}{!}{\includegraphics{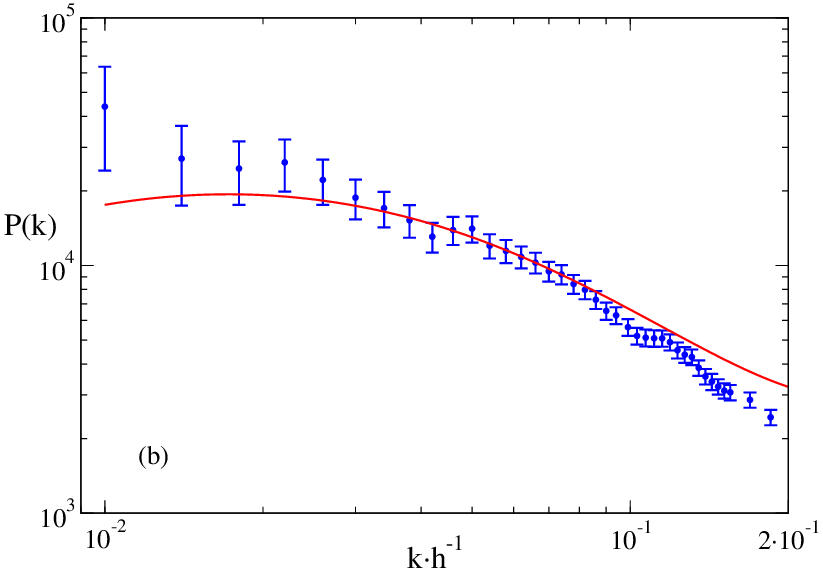}}}\\

   \\
   \\
   \mbox{\resizebox*{0.45\textwidth}{!}{\includegraphics{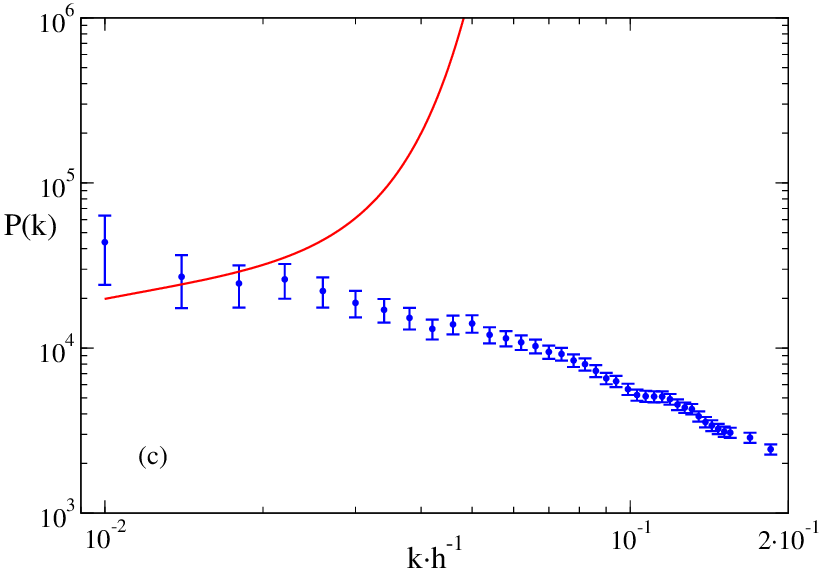}}}
\caption{ As in Fig\,\protect\ref{fig3} (running cosmological
constant model), but for fixed $\Omega_{B}^0 = 0.04$,
$\Omega_{DM}^0 = 0.21$ and $\Omega_{\Lambda}^0 = 0.75$ and for the
following (negative) values of $\nu$: (a) $- 10^{-8}$, (b) $-
10^{-6}$, and (c) $- 10^{-4}$, together with the LSS data from the
2dfFGRS\,\cite{cole}. } \label{fig4}}
%%%%%%%%%%%%%%%%%%%%%%%%%%%%%%%%%%%%%%%%%%%%%%%%%%%%%%%%%%%%%%%%%%%%%%%%%%

%%%%%%%%%%%%%%%%%%%%%%%%%%%%%%%%%%%%%%%%%%%%%%%%%%%%%%%%%%%%%%%%%%%
\FIGURE[t]{ \centering
\resizebox{0.87\textwidth}{!}{\includegraphics{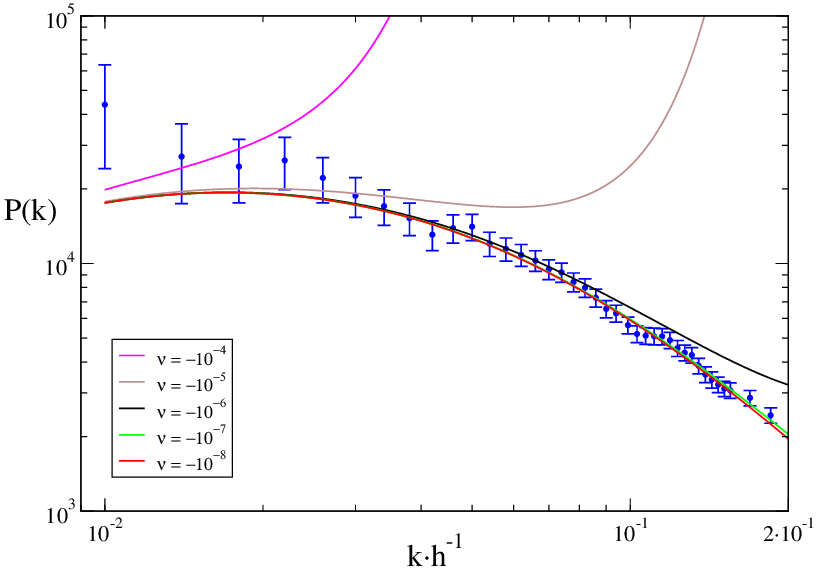}}
\caption{As in Fig\,\protect\ref{fig3} (running cosmological
constant model), but for fixed $\Omega_{B}^0 = 0.04$,
$\Omega_{DM}^0 = 0.21$ and $\Omega_{\Lambda}^0 = 0.75$ with
different negative values of $\nu$ indicated in the figure,
together with the LSS data from the 2dfFGRS\,\cite{cole}.  The
curves corresponding to the last two values of $\nu$ in the list
almost coincide.}\label{fig5} }

%%%%%%%%%%%%%%%%%%%%%%%%%%%%%%%%%%%%%%%%%%%%%%%%%%%%%%%%%%%%%%%%%%%%
%%%%%%%%%%%%%%%%%%%%%%%%%%%%%%%%%%%%%%%%%%%%%%%%%%%%%%

%%%%%%%%%%%%%%%%%%%%%%%%%%%%%%%%%%%%%%%%%%%%%%%%%%%%%%
\section{Discussion and conclusions}
\label{sect:Conclusion}

The possibility that the cosmological constant $\Lambda$ can be a
running parameter within QFT in curved space-time was investigated
extensively in \,\cite{nova}. The phenomenological implications of
these renormalization group (RG) cosmologies on supernovae
observations have been amply addressed in different works preceding
this one \,\cite{RGCC,IRGA03,babic,CCfit,AHEP03,Gruni,SS1,SolSte}.
Here we have extended the analysis of phenomenological implications
to the very relevant issue of structure formation. We have studied
in detail the power spectrum of matter density perturbations for the
FLRW type of cosmological model with running $\Lambda$. This model
is characterized by a fundamental parameter $\nu$ associated to the
running of $\Lambda$. At the same time this parameter acts, in the
present realization of the model, as a coupling between the vacuum
energy and matter (mainly CDM). There are other implementations of
this framework where $\nu$ does not lead to this kind of coupling,
see \cite{Gruni}. However, in the present work we have just focused
on the original RG model\,\cite{RGCC,CCfit}, where this coupling is
present and, therefore, the cosmological term can decay into matter
and vice versa. Then, under the simplifying assumptions of
adiabaticity (i.e. of having no perturbations associated to entropy
exchange) and of having no anisotropic stress contributions, we have
derived the complete coupled set of matter density and metric
perturbations for this model. The RG running of $\Lambda$ is based
on the assumption of a standard quadratic decoupling law at low
energies\,\cite{RGCC,CCfit} and we have taken the simplest covariant
form of this equation -- see (\ref{A and B}) -- as our ansatz to
derive the perturbations. This phenomenological input leads to a
cosmological model with the $H^2$-dependence of the vacuum energy
density\,\cite{nova}. It is important to emphasize that this does
not mean that $\Lambda\sim H^2$, but rather that the
(renormalization group) variation of $\Lambda$ satisfies
$\delta\Lambda\sim H^2$, and hence $\Lambda=a+b\,H^2$ -- see
Eq.\,(\ref{Friedmann2}). The effect produced by the running depends
on a single parameter $\nu$, which characterizes the particle mass
spectrum of the quantum theory below the Planck scale. For $\nu=0$
we retrieve the standard $\Lambda$CDM model with time-independent
cosmological constant, while the values around the ``canonical''
value $\nu=\nu_0\sim 10^{-2}$ correspond to having effective
particle contributions of order of the Planck mass.

The numerical analysis of the perturbations proves to be in
qualitative agreement with the results of the previous estimate in
Ref.\,\cite{OphPel}, based on restraining the amplification of the
density fluctuating spectrum at the recombination era caused by
the vacuum decay into CDM.  In both cases (viz. the direct
calculation and the previous estimate) it is found that the
``canonical'' value of $\nu$ leads to strong deviations from the
observational data. For positive values of $\nu$ in the range
$10^{-6}\leqslant\nu\leqslant 10^{-5}$, corresponding
approximately to the GUT particle spectrum ($M_X \sim
(10^{-2}-10^{-3})\,M_P$), we meet relatively small deviations of
the theoretical curve from the $\nu=0$ case. The deviations are
higher the larger is $|\nu|$. Let us recall that the upper bound
on $\nu$ obtained from the phenomenological estimate
of\,\cite{OphPel} is $10^{-3}$ and applies only for $\nu\geqslant
0$. This bound is not very far away from our own most conservative
result $|\nu|<10^{-4}$, although in our case it applies for both
signs of $\nu$. Moreover, the approximate bound is much weaker
than the one defined by the so-called ``safest $\nu$ range'' --
cf. Eq.\, (\ref{nurange}). However, as we already emphasized, we
cannot definitely exclude the $\nu>0$ interval $10^{-6}-10^{-4}$
(and specially the subinterval $10^{-6}-10^{-5}$). Therefore, we
may assert, in the most possible conservative way, that our direct
bounds on $\nu> 0$ are between one to two orders of magnitude more
restrictive than those of \cite{OphPel}, but of course they have
been derived in a much more rigorous way.  Let us also remark
that our exclusion of the possibility that $\nu$ could be of the
order of magnitude of the ``canonical'' value $\nu_0\sim 10^{-2}$
is at variance (by at least two orders of magnitude) with the
estimates obtained from the approximate methods used in
references \cite{WangMeng,AlcLim}\,\footnote{The results of these
references were presented in terms of a parameter $\epsilon$, and
they concluded that values of $\epsilon$ at the level of
$10^{-1}$ are allowed. However, $\epsilon$ is nothing but $3$
times the parameter $\nu$ originally defined in \cite{RGCC}.}.

It is also interesting to compare our results with those of
Ref.\,\cite{pavon}, where a detailed study of matter density
perturbations in interacting quintessence models is addressed. These
authors perform their study also under the assumption of an
adiabatic regime and with no contributions from anisotropic stress.
In this kind of modified quintessence models, first proposed in
\,\cite{Amendola}, one couples matter to quintessence by means of a
source function $Q$ which depends on a free parameter, called $c^2$
in Ref.\,\cite{pavon}. This is the only free parameter in that
model, similarly as is $\nu$ in our own framework under the
specified common set of assumptions. In the models of interactive
quintessence, the dark energy (DE) can also decay into matter
(parallel to our case where $\Lambda$ can also decay into matter)
and as a result the amount of DE in the past is larger than in
ordinary quintessence models with self-conserved DE. This
corresponds, in our framework, to have $\nu>0$ because then
$\Lambda$ becomes larger in the past. The authors of \,\cite{pavon}
find that the effect of the $c^2$-coupling is to introduce a damping
of the power spectrum of matter perturbations at low scales; that is
to say, the growth of density perturbations for large wave numbers
becomes smaller as compared to quintessence models with uncoupled DE
to matter. In our RG framework, we observed a similar depletion of
the spectrum at low scales for $\nu>0$ as compared to $\nu=0$ (the
standard $\Lambda$CDM model) -- cf. sections \ref{sect:initialcond}
and \ref{sect:powerspectrum}. Moreover, we have found numerical
bounds on $\nu$ that are quantitatively similar, in fact a bit more
stringent than those on $c^2$ obtained in the model of
\,\cite{pavon}. These authors found that their power spectrum is
consistent with the 2dFGRS data provided $c^2\leq 10^{-3}$, and we
find that the largest allowed value of $\nu$ by the same LSS data is
of order $10^{-4}$. The basic agreement among the different analyses
suggest that the loss of power at low scales could be a general
feature of dynamical models in which the dark energy decays into
matter during the cosmic evolution. The latter situation is always
the case in Ref. \,\cite{pavon} because $c^2>0$, and it is also our
case for $\nu>0$. However, in our RG framework, $\nu$ can have any
sign (depending on the dominance of bosons or fermions in the loop
contributions to the running of $\Lambda$) and so we have an
opportunity to explore also the case $\nu<0$. For this sign of
$\nu$, the running $\Lambda$ becomes progressively smaller in the
past and reaches eventually large negative values (see the detailed
evolution plots of Ref.\,\cite{CCfit}). As a result we may expect an
overproduction of structure associated to the fact that a large
negative $\Lambda$ increases the gravitational collapse. Since,
however, this is not observed (cf. the plots of section
\ref{sect:powerspectrum} detecting an excess of power at low scales
for $\nu<0$), this may provide a physical explanation of why
negative values of $\nu$ are significantly more restrained. While
the safest and most stringent range of values of $\nu$ is expressed
in a nutshell in Eq.\,(\ref{nurange}), we cannot exclude that
$|\nu|$ could be higher (mainly in the $\nu>0$ case) once it will be
possible to combine the analysis of the matter power spectrum with
the corresponding CMB analysis. At the same time, we should recall
that the bounds obtained on $\nu$ in our case, and on $c^2$ in
\,\cite{pavon}, could suffer some renormalization if we would
abandon the set of simplifying hypotheses that have been made.
However, our study shows at least the kind of impact to be expected
on the matter power spectrum predicted by cosmological models with
running $\Lambda$ in interaction with matter. Further investigations
would be needed to assess the effect in a more thorough way, but
they lie beyond the scope of the present work.

Let us also finally remark that the bounds we have placed on the RG
parameter $\nu$, within our set of assumptions, can be interpreted
on purely phenomenological grounds, namely as bounds on a parameter
that gauges the possible degree of interaction between CDM and the
vacuum energy, even if no fundamental RG model is invoked. In
principle, the phenomenological implications of a non-vanishing
$\nu$ could be potentially detectable in the future supernovae
observations, either through a direct measure of the running of
$\Lambda$ \,\cite{CCfit,AHEP03} or (more efficiently in practice)
through the possibility to observe the redshift evolution of the
effective equation of state parameter associated to the variable CC
model \,{\cite{SS1,SolSte}}. However, the very study of the
evolution of the density and metric perturbations, which has led us
to firmly exclude the range $|\nu|>10^{-4}$, already demonstrates
the difficulty of the previous methods and at the same time
emphasizes the great advantage of the density perturbations
approach. Indeed, the strong sensitivity of the matter power
spectrum to a variable cosmological term interacting with the CDM
shows that this is perhaps the ideal observable to look at.
Therefore, future data on LSS could eventually provide the clue to
unravel an interaction (even if very small) between matter
(essentially CDM) and vacuum, and at the same time it may hint at
the possible RG physics behind it. Let us conclude by mentioning
that it should be very interesting to expand our analysis to the
aforementioned second model of running CC \cite{Gruni}. In this
alternate model the cosmological term is also evolving in time and
redshift, but there is no energy exchange between matter and vacuum
and we may expect much more freedom in choosing the value of $\nu$.
We leave this study for a future publication. \vspace{0.3cm}

%%%%%%%%%%%%%%%%%%%%%%%%%%%%%%%%%%%%%%%%%%%%%%%%%%%%%%%%%%%%%%%%%%
{\bf Acknowledgments:} We are grateful to J\'er\^ome Martin for
useful discussions, to Hrvoje \v{S}tefan\v{c}i\'{c} for checking
our numerical code and for interesting discussions, and finally
to Javier Grande for his aid in improving the presentation of our
plots and also for useful discussions. The work of J.C.F has been
supported in part by CNPq (Brazil) and by CAPES/COFECUB
Brazilian-French scientific cooperation. The work of I.Sh. has
been supported in part by the grants from CNPq (Brazil), FAPEMIG
(Minas Gerais, Brazil) and ICTP (Italy). The work of J.S. has been
partially supported by MECYT and FEDER under project
2004-04582-C02-01, and also by DURSI Generalitat de Catalunya
under project 2005SGR00564.

%\newpage
%%%%%%%%%%%%%%%%%%%%%%%%%%%%%%%%%%%%%%%%%%%%%%%%%%%%%%

%%%%%%%%%%%%%%%%%%%%%%%%%%%%%%%%%%%%%%%%%%%%%%%%%%%%%%%%%%%%%%%%
\end{document}